\documentclass[journal,draftcls,onecolumn,11pt]{IEEEtran}
\usepackage{arxiv}
\usepackage[utf8]{inputenc} 
\usepackage[T1]{fontenc}    
\usepackage{hyperref}       
\usepackage{url}            
\usepackage{booktabs}       
\usepackage{amsfonts}       
\usepackage{nicefrac}       
\usepackage{microtype}      
\usepackage{lipsum}
\usepackage{amsmath}
\usepackage[ruled,vlined,linesnumbered,lined]{algorithm2e}
\newcommand{\norm}[1]{\left\lVert#1\right\rVert}
\newtheorem{theorem}{Theorem}

\newtheorem{example}[theorem]{Example}

\usepackage{tikz}
\usetikzlibrary{arrows,shapes,chains,matrix,positioning,scopes,patterns,calc,decorations.pathmorphing,shadows}
\usepackage{subcaption}
\usepackage{pgfplots}
\pgfplotsset{
        compat=1.14,
    }
\usepackage[underline=true,rounded corners=false]{pgf-umlsd}
\tikzstyle{cnode}=[circle,minimum size=0.75cm,draw]
\tikzstyle{cgnode}=[circle,draw]
\tikzstyle{crnode}=[circle,draw]
\tikzstyle{conode}=[rectangle,draw,fill=black!15]
\tikzstyle{cpnode}=[circle,draw]
\tikzstyle{rnode}=[rectangle,draw,outer sep=0pt]
\tikzstyle{fnode}=[rectangle,draw,minimum height=0.6cm,minimum width=1.1cm,fill=red!75,text centered]
\tikzstyle{prnode}=[rectangle,minimum height=0.6cm,minimum width=1.1cm,text centered,draw]
\tikzstyle{prnodebig}=[rectangle,draw,rounded corners,minimum width=8cm,minimum height=4cm,text centered,outer sep=0pt]
\tikzstyle{prnodesimple}=[rectangle,draw,text width=11em,text centered,outer sep=0pt]
\tikzstyle{bigsnake}=[snake=snake,segment amplitude=4mm, segment length=4mm, line after snake=5mm]
\tikzstyle{smallsnake}=[snake=snake,segment amplitude=0.7mm, segment length=4mm, line after snake=3mm]
\pgfplotsset{compat=1.14}

\title{Factored LT and Factored Raptor Codes for Large-Scale Distributed Matrix Multiplication}

\begin{document}

\author{\begin{tabular}{c} Asit Kumar Pradhan, 
Anoosheh Heidarzadeh,
Krishna R. Narayanan
\end{tabular} \\
Department of Electrical and Computer Engineering \\
Texas A\&M University}
\maketitle
\begin{abstract}
We propose two coding schemes for distributed matrix multiplication in the presence of stragglers. 
These coding schemes are adaptations of LT codes and Raptor codes to distributed matrix multiplication
and are termed \emph{factored LT (FLT) codes} and \emph{factored Raptor (FR) codes}.
Empirically, we show that FLT codes have near-optimal recovery thresholds when the number of worker nodes is very large, and that FR codes have excellent recovery thresholds while the number of worker nodes is moderately large. 
FLT and FR codes have better recovery thresholds when compared to Product codes and they are expected to have better numerical stability when compared to Polynomial codes, while they can also be decoded with a low-complexity decoding algorithm.
\end{abstract}

\section{Introduction}
We consider a matrix multiplication problem where the aim is to compute $\mathbf{C}=\mathbf{A}^\mathsf{T}\mathbf{B}$ given two input matrices $\mathbf{A} \in \mathbb{R}^{s \times r}$ and $\mathbf{B} \in \mathbb{R}^{s \times t}$ for some integers $r, s$ and $t$. 
Many applications in optimization and machine learning require multiplications of large matrices of dimension of the order of $10^5 \times 10^5$. 
These large-scale matrix multiplications cannot be carried out in a single machine mainly due to low-latency requirement in many applications. The low-latency requirement can be met by dividing input matrices $\mathbf{A}$ and $\mathbf{B}$  into $m$ and $n$ sub-matrices, respectively and then distributing the tasks of computing product of these sub-matrices among all worker nodes. A master node collects the partial results from the worker nodes and aggregates them appropriately to obtain $\mathbf{C}$. Since the master depends on all the worker nodes to compute $\mathbf{C}$, a few slow worker nodes, referred to as \emph{stragglers}, can increase the computational delay. 
For a given scheme, we define the recovery threshold as the minimum number of workers that the master node needs to wait for to compute $\mathbf{C}$.

In \cite{speeding_up}, Lee \emph{et al.}~proposed a scheme, referred to as \emph{1-D maximum distance separable (MDS) codes}, to mitigate the effect of stragglers for the case $n=1$, in which the sub-matrices of the matrix $\textbf{A}$ are encoded using an $(\tilde{m},m)$ MDS code and the task of computing $\mathbf{A}_i^\mathsf{T}\mathbf{B}$, for $i \in [\tilde{m}]$, is distributed among $\tilde{m}$ worker nodes.
This scheme can be extended to matrix-matrix multiplication in a natural way by considering each column of $\mathbf{B}$ as a vector; however, as shown in \cite{yu2018straggler}, the resulting threshold will not be optimal. 
In \cite{yu2018straggler}, Yu \emph{et al.}~proposed a coding scheme, referred to as \emph{Polynomial codes}, which divides both $\mathbf{A}$ and $\mathbf{B}$ into sub-matrices. In \cite{yu2018straggler}, the authors have also shown that the minimum achievable recovery threshold is $K=mn$ and Polynomial codes have optimal recovery threshold. 
The main drawback of Polynomial codes is that the decoding process requires interpolating polynomials of degree-$K$, which is equivalent to inverting a $K \times K$ Vandermonde matrix. 
It is well known that inverting Vandermonde matrices is highly numerically unstable even for moderate values of $K$.  

In \cite{highdim}, Lee \emph{et al.}~proposed a coding scheme using Product codes, which encodes both the input matrices $\mathbf{A}$ and $\mathbf{B}$ using an $(\tilde{m},m)$ MDS code unlike the scheme in \cite{speeding_up}.  
Product codes can be implemented using several component codes such as Polynomial codes \cite{yu2018polycode}, OrthoPoly codes \cite{fahim2019}, random Khatri-Rao-Product (RKRP) codes \cite{adarsh2019RKRP}, etc.
Product codes are generally better than Polynomial codes in terms of  numerical stability.
For example, decoding of a Product code of dimension $K$ built from Polynomial component codes of dimension $\sqrt{K}$ requires the inversion of $\sqrt{K} \times \sqrt{K}$ Vandermonde matrices, whereas decoding of a Polynomial code of dimension $K$ would require 
inversion of a $K \times K$ Vandermonde matrix.
The main drawback of Product codes is that their recovery threshold is not optimal unlike Polynomial codes. In \cite{ddimprod}, Baharav \emph{et al.}~proposed a scheme, referred to as \emph{$d$-dimensional Product codes} for matrix multiplication, by spreading component matrices of $\mathbf{A}$ and $\mathbf{B}$ over $d/2$ dimensions each, and encoding them using a $d$-dimensional Product code. 
While $d$-dimensional Product codes can perform better than $2$-dimensional Product codes for certain
regimes of $N$ and $K$, the recovery threshold of $d$-dimensional Product codes is still not optimal when the number of stragglers is linear in $N$.  Moreover, $(N,K)$ $d$-dimensional Product codes of rate $R$ are built upon $(N^{1/d},K^{1/d})$ MDS component codes of rate $R^{1/d}$. Unless $N$ is very large, it is impractical to use large values of $d$ since the set of possible MDS component codes becomes trivial for large $d$.

In~\cite{Mallick}, Mallick \emph{et al.}~proposed a scheme for matrix-vector multiplication using Luby Transform (LT) codes~\cite{LTcodes}, a type of rateless codes, which has many desired properties: (i) Numerical stability; (ii) Linear decoding complexity, and (iii) Recovery threshold of $P(1 - \alpha)$, where $\alpha \in [0,1)$ is the fraction of worker nodes that are stragglers. However, their encoding scheme is not directly extendable to the matrix-matrix multiplication problem. In \cite{Shroff}, Wang \emph{et al.}~proposed the use of LT codes for distributed matrix multiplication which has all the three desired properties; however, for this scheme, both the communication and the computation at each worker node are substantially more expensive than those of Polynomial codes and Product codes.

In this paper, we propose two novel encoding schemes. The first one is based on LT codes, referred to as \emph{factored LT (FLT) codes}, which is better in terms of numerical stability as well as decoding complexity  when compared to Polynomial codes. 
In particular, the decoding complexity of FLT codes is $\mathcal{O}(rt \log K)$, whereas the decoding complexity of Polynomial code is $\mathcal{O}(rt\log^2K \log\log K)$. 
As in the case of LT codes for erasure channels, the performance of FLT codes can be improved for finite lengths, particularly for moderate values of $K$.
For this regime, we propose a Raptor code based scheme, referred to as \emph{factored Raptor (FR) codes}, which performs well when $K$ is moderately large. Simulation results show that the recovery thresholds of FR codes are better than those of Product codes. For example, a $(10000,6400)$ FR code has recovery threshold $7060$, whereas the recovery threshold of an $(21,18)\times(22,19) \times (22,19)$ Product code is $7275$. 

\section{Notation}
We use boldface capital letters for matrices and underlined variables to represent vectors. We denote the $(i,j)$th element of the matrix $\mathbf{A}$ by $A_{ij}$. We denote the set of integers from $1$ to $i$ by $[i]$. We represent the $i$th row and $j$th column of the matrix $\mathbf{A}$ by $\mathbf{A}_{i,:}$ and $\mathbf{A}_{:,j}$, respectively. We assume that vectors without transposes are column vectors unless stated otherwise. 
We denote the cardinality of a set $\mathcal{S}$ by $|\mathcal{S}|$.  
\section{System Model}
\label{sec:sysmodel}
We consider a system where there is one master node and $P$ worker nodes. The goal of the master node is to compute $\mathbf{C}=\mathbf{A}^\mathsf{T}\mathbf{B}$ in a distributed fashion using $P$ worker nodes. Worker nodes can only communicate with the master node and cannot communicate among themselves. To distribute the computation among workers, $\mathbf{A}$ and $\mathbf{B}$ are divided along columns to sub matrices of size $s \times \frac{r}{m}$ and $s \times \frac{t}{n}$, respectively, as shown below.
\begin{equation}
    \mathbf{A}=[\mathbf{A}_1,\mathbf{A}_2, \cdots, \mathbf{A}_m] \text{ \; and \; } \mathbf{B}=[ \mathbf{B}_1, \mathbf{B}_2, \cdots \mathbf{B}_n].
\end{equation}
Alternatively, the output matrix $\mathbf{C}$ can be written in terms of the components of $\mathbf{A}$ and $\mathbf{B}$ as follows:
    \[
\mathbf{C} = \begin{bmatrix} 
    \mathbf{A}_{1}^\mathsf{T}\mathbf{B}_1 & \mathbf{A}_{1}^\mathsf{T}\mathbf{B}_2 & \dots & \mathbf{A}_{1}^\mathsf{T}\mathbf{B}_n \\
    \mathbf{A}_{2}^\mathsf{T}\mathbf{B}_1 & \mathbf{A}_{2}^\mathsf{T}\mathbf{B}_1 & \dots & \mathbf{A}_{2}^\mathsf{T}\mathbf{B}_n \\
    \vdots & \vdots & \ddots & \vdots \\
    \mathbf{A}_{m}^\mathsf{T}\mathbf{B}_1 & \mathbf{A}_{m}^\mathsf{T}\mathbf{B}_2 &\dots & \mathbf{A}_{m}^\mathsf{T}\mathbf{B}_n
    \end{bmatrix}.
\]
Then computing $\mathbf{C}$ is equivalent to computing $mn$ blocks of $\mathbf{A}_{i}^{\mathsf{T}}\mathbf{B}_j$ for $ i \in [m]$ and $j \in [n]$. For each $p\in [P]$, let $\mathcal{I}^{p}_{\mathbf{A}}$ and $\mathcal{I}^{p}_{\mathbf{B}}$ be two subsets of $[m]$ and $[n]$, respectively. For each $p\in [P]$, we define
     \begin{equation}
     \label{eq:3}
        (\mathbf{\widetilde{A}}^p)^\mathsf{T}=\sum_{i \in [m]}a_i^p\mathbf{A}_i^\mathsf{T}, \quad \mathbf{\widetilde{B}}^p=\sum_{j \in [n]}b_j^p\mathbf{B}_j,
    \end{equation}
    where   $a_i^p$ for $i \in \mathcal{I}_{\mathbf{A}}^p$ and $b_j^p$ for $j \in \mathcal{I}_{\mathbf{B}}^p$ are randomly sampled from a Gaussian distribution with mean zero and variance one; and we choose $a_i^p=b_j^p=0$ for $i \notin \mathcal{I}_{\mathbf{A}}^p$
and $j \notin \mathcal{I}_{\mathbf{B}}^p$.
    The master node computes $(\mathbf{\widetilde{A}}^p)^\mathsf{T}$ and $\mathbf{\widetilde{B}}^p$ and sends them to the worker node $p$.
   The worker node $p$ computes $\mathbf{\widetilde{C}}^p=(\mathbf{\widetilde{A}}^p)^\mathsf{T}\mathbf{\widetilde{B}}^p$ and returns $\mathbf{\widetilde{C}}^p$ to the master node. In literature, $  \mathbf{ \widetilde{ C}}^\mathsf{T}=[\mathbf{\widetilde{C}}^1,\mathbf{\widetilde{C}}^2,\cdots \mathbf{\widetilde{C}}^P]$ is referred to as the \emph{encoding 
   function}. The master node collects $\mathbf{\widetilde{C}}^p$'s from a subset of worker nodes, referred to as \emph{non-stragglers}, and attempts to recover the matrix $\mathbf{C}$ from the results of the non-straggling worker nodes using a decoding function $f$.
    For a given encoding function $\mathbf{\widetilde{C}}$ and decoding function $f$, the recovery threshold is defined as the minimum integer $N$ such that the master node can recover the matrix $\mathbf{C}$ from the results of any $N$ (non-straggling) worker nodes. The goal is to design an encoding function and a decoding function so as to minimize the recovery threshold. 

\section{Proposed Coding Schemes}
\subsection{LT-Coded Distributed Matrix Multiplication}
In this section, we propose an LT code based scheme, termed \emph{factored LT (FLT) codes}, to mitigate the effect of straggling workers in the computation of $\mathbf{C}=\mathbf{A}^\mathsf{T}\mathbf{B}$. We briefly describe encoding and decoding of the proposed scheme below.
\subsubsection{Encoding}
\label{sec:LTcodes}
LT codes, introduced by Luby, are a class of rateless erasure codes that can be used to generate a (potentially infinite) sequence of output symbols from $K$ source symbols. 
The number of source symbols involved in generating an output symbol is referred to as the \emph{degree of the output symbol}. The output symbols follow a degree distribution $\Omega(x)$. In \cite{LTcodes}, the authors have shown that in the case of the single user erasure channel, the source symbols can be recovered from any $N=K(1+\epsilon)$ output symbols with high probability, where $\epsilon$ is the overhead. 
In \cite{LTcodes}, it was shown that for some proper choice of degree distribution $\Omega(x)$, the overhead $\epsilon$ vanishes as $K$ grows unbounded.
To apply LT codes to the task of matrix-matrix multiplication, we treat $\mathbf{A}_i^\mathsf{T}\mathbf{B}_j$'s for $i \in [m]$ and $j \in [n]$ as source symbols, where $\mathbf{A}_i$'s and $\mathbf{B}_j$'s are component matrices of $\mathbf{A}$ and $\mathbf{B}$, respectively.
As described in Section~\ref{sec:sysmodel}, the matrix-matrix multiplication problem requires each output symbol to be the product of $\mathbf{\widetilde{A}}$ and $\mathbf{\widetilde{B}}$, where $\mathbf{\widetilde{A}}$ and $\mathbf{\widetilde{B}}$ are sum of randomly chosen chunks of $\mathbf{A}$ and $\mathbf{B}$, respectively. 
Therefore, unlike in the case of the single user erasure channel, a degree-$d$ output symbol cannot be generated from any $d$ source symbols for the matrix-matrix multiplication problem. 
So the encoding of LT codes cannot be directly applied to the matrix-matrix multiplication problem.

 Let $\Omega(x)=\sum_{i=1}^K \Omega_ix^i$ be a degree distribution, where $\Omega_i=0$ for all prime $i>\max(m,n)$. In the encoding process of the FLT codes,  the master node does the following for each worker $p \in [P]$:
\begin{itemize}
\item[1.] Randomly sample a degree $d$ by sampling from $\Omega(x)$.
\item[2.] Randomly choose a  divisor, denoted by $d_1$, of $d$ in such a way that $d_1 \leq m$ and $\frac{d}{d_1} \leq n$. Denote $\frac{d}{d_1}$ by $d_2$.
    \item[3.] Let $\mathcal{S}_m$ ($\mathcal{S}_n$) be the collection of all subsets of the set $[m]$ ($[n]$) that are of size $d_1$ ($d_2$). Choose $\mathcal{I}_{\mathbf{A}}^p$ and $\mathcal{I}_{\mathbf{B}}^p$ uniformly randomly from the set $\mathcal{S}_m$ and $\mathcal{S}_n$, respectively. 
    \item[4.]
    Compute $(\mathbf{\widetilde{A}}^p)^\mathsf{T}$ and $\mathbf{\widetilde{B}}^p$ as in \eqref{eq:3}, and send them to the worker node $p$.
\end{itemize}
Each worker node $p$ computes the output symbol $\mathbf{\widetilde{C}}^p = (\mathbf{\widetilde{A}}^p)^\mathsf{T}\mathbf{\widetilde{B}}^p$, and returns $\mathbf{\widetilde{C}}^p$ to the master node. The collection of $\mathbf{\widetilde{C}}^p$'s for $p \in [P]$ form a codeword of the FLT code.

\begin{example}
\label{ex:Ltencoding}
Let $m = 3$, $n=3$ and $\Omega(x) = 0.2 x + 0.7 x^2 + 0.1 x^4$. 
Consider the generation of an encoded symbol at the worker node $p$. 
We randomly sample a $d$ from $\Omega(x)$ and say we obtain $d=4$.
Let $d_1=2$, which implies that $d_2=2$. Therefore, $\mathcal{S}_{\mathbf{\widetilde{A}}}^{p}=\{\{1,2\},\{1,3\},\{2,3\}\}$ and $\mathcal{S}_{\mathbf{\widetilde{A}}}^{p}=\{\{1,2\},\{1,3\},\{2,3\}\}$. 
We choose $\mathcal{I}_{\widetilde{A}}^{p}$ and $\mathcal{I}_{\widetilde{B}}^{p}$ uniformly at random from 
$\mathcal{S}_{\mathbf{\widetilde{\textbf{B}}}}^{p}$,  respectively and let
let $\mathcal{I}_{\widetilde{A}}^{p}=\{2,3\}$ and $\mathcal{I}_{\widetilde{B}}^{p}=\{1,3\}$. 
In this example, for ease of exposition, we choose $a_i^p$'s and $b_j^s$'s as follows:
\begin{equation*}
        a_i^p= \begin{cases}
        1, \text{ if } i \in \mathcal{I}_{\mathbf{A}}^p \\
        0, \text{ Otherwise }
        \end{cases}, \quad
        b_j^p= \begin{cases}
        1, \text{ if } j \in \mathcal{I}_{\mathbf{B}}^p \\
        0, \text{ Otherwise},
        \end{cases}
    \end{equation*} 
    
The master node computes $(\mathbf{\widetilde{A}}^\mathsf{T})^{p}=\mathbf{A}_2^{\mathsf{T}}+\mathbf{A}_3^{\mathsf{T}}$, $\mathbf{\widetilde{B}}^{p}=\mathbf{B}_1+\mathbf{B}_3$ and sends them to the worker node $p$
which computes 
$\mathbf{\widetilde{C}}^{p}= \mathbf{\widetilde{A}}^{\mathsf{T}^{p}} \mathbf{\widetilde{B}}^{p} =  (\mathbf{A}_2^{\mathsf{T}}+\mathbf{A}_3^{\mathsf{T}})(\mathbf{B}_1+\mathbf{B}_3)$ and sends it to the master node. 
Notice that $\mathbf{\widetilde{C}}^{p} = \mathbf{A}_2^{\mathsf{T}} \mathbf{B}_1 + \mathbf{A}_2^{\mathsf{T}} \mathbf{B}_3 + \mathbf{A}_3^{\mathsf{T}} \mathbf{B}_1 + \mathbf{A}_3^{\mathsf{T}} \mathbf{B}_3$ is a linear combination of 4 source symbols. 
The encoding process is illustrated in Fig.~\ref{fig:encoding_deg4_node}.
\end{example}
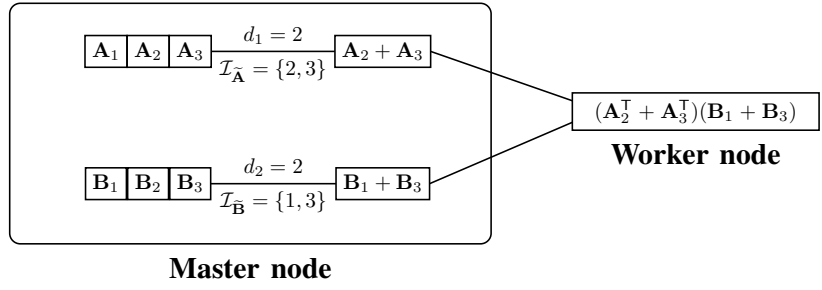
\begin{figure}[h]
    \centering
    \begin{tikzpicture}[every node/.style={scale=0.8}]
\begin{scope}[node distance=1.2cm,>=angle 90,semithick]
\node[rnode] (a1) {$\mathbf{A}_1$};
\node[rnode] (a2)[right of=a1,xshift=-0.5cm] {$\mathbf{A}_2$};
\node[rnode] (a3)[right of=a2,xshift=-0.5cm] {$\mathbf{A}_3$};
\node[prnodebig] (M)[below of=a3,xshift=1cm] {};
\node [below of=M,yshift=-1.2 cm,scale=1.4] {\textbf{Master node}};
\node[rnode] (b1)[below of=a1,yshift=-1cm] {$\mathbf{B}_1$};
\node[rnode] (b2)[right of=b1,xshift=-0.5cm] {$\mathbf{B}_2$};
\node[rnode] (b3)[right of=b2,xshift=-0.5cm] {$\mathbf{B}_3$};
\node[rnode] (a22)[right of=a3,xshift=2cm] {$\mathbf{A}_2+\mathbf{A}_3$};
\node[rnode] (b22)[right of=b3,xshift=2cm] {$\mathbf{B}_1+\mathbf{B}_3$};
\draw (a3) -- node[above]  {$d_1=2$} node[below] {$\mathcal{I}_{\mathbf{\widetilde{A}}}=\{2,3\}$} (a22);
\draw (b3) -- node[above]  {$d_2=2$} node[below] {$\mathcal{I}_{\mathbf{\widetilde{B}}}=\{1,3\}$} (b22);
\node[prnodesimple] (w)[right of=a22,xshift=4cm,yshift=-1cm] {$(\mathbf{A}_2^{\mathsf{T}}+\mathbf{A}_3^{\mathsf{T}})(\mathbf{B}_1+\mathbf{B}_3)$};
\node [below of=w,yshift=0.5 cm,scale=1.4] {\textbf{Worker node}};
\draw (a22.0) -- (w.175);
\draw (b22.0) -- (w.185);
\end{scope}    
\end{tikzpicture}
    \caption{Encoding of a degree-$4$ node}
    \label{fig:encoding_deg4_node}
\end{figure}
\subsubsection{Decoding}
\label{sec:LTcodes_decoding}
 Without loss of generality, suppose that the master node collects results from the first $N$ workers with $N \leq P$. Given the above encoding scheme, we have
\[
\begin{bmatrix}
    \mathbf{\widetilde{C}}^1      \\
    \mathbf{\widetilde{C}}^2   \\
      \vdots \\
     \mathbf{\widetilde{C}}^N   
\end{bmatrix}
= 
\begin{underbrace}{
\begin{bmatrix} 
     a_{1}^1b_1^1 & a_{1}^1b_2^1 & \dots & a_{m}^1b_n^1 \\
    a_{1}^2b_1^2 & a_{1}^2b_1^2 & \dots & a_{m}^2b_n^2 \\
    \vdots & \vdots & \ddots & \vdots \\
    a_{1}^Nb_1^N & a_{1}^Nb_2^N &\dots & a_{m}^Nb_n^N \\
\end{bmatrix}
}_{\mathbf{\mathlarger{M}}}
\end{underbrace}
\begin{bmatrix}
    \mathbf{A}_1^\mathsf{T}\mathbf{B}_1     \\
    \mathbf{A}_1^\mathsf{T}\mathbf{B}_2   \\
      \vdots \\
     \mathbf{A}_m^\mathsf{T}\mathbf{B}_n   
\end{bmatrix}
    \]
We use $\mathbf{M} \in \mathbb{R}^{N\times mn}$ to represent the above coefficient matrix. To guarantee decodability, the master node should collect results from enough number of workers such that the coefficient matrix $\mathbf{M}$ has full column rank. Then the master node goes through a peeling decoding process, which can be described using a bipartite graph. The bipartite graph can be constructed with one partition being the set of source symbols $\mathbf{A}_i^\mathsf{T}\mathbf{B}_j$, $i \in [m]$ and $j \in [n]$, and the other partition being the set of output symbol (computation tasks) $\mathbf{\widetilde{C}}^p, \ p \in [N]$. There is an edge between the node $\mathbf{A}_i^\mathsf{T}\mathbf{B}_j$ and $\mathbf{\widetilde{C}}^p$ if $\mathbf{A}_i^\mathsf{T}\mathbf{B}_j$ is involved in the computation of $\mathbf{\widetilde{C}}^p$.  The peeling decoding process works in iterations. In each iteration, the master node searches for a right node of degree 1. If such a node cannot be found, the decoding process terminates. Otherwise, the master node recovers the (unique) left node connected to the found degree-1 right node, and removes all edges adjacent to the recovered left node. The master node runs this process iteratively until all left nodes are recovered or no degree-1 right node can be found. An early termination of the decoding process (before all left nodes are recovered) yields a decoding failure; otherwise, the decoding process is dubbed successful.

\subsection{Raptor-Coded Distributed Matrix Multiplication} 
In the FLT coding scheme presented in section~\ref{sec:LTcodes},  the output symbols are generated by taking a linear combination of randomly chosen source symbols. Therefore, when the number of encoded symbols are close to the number of source symbols, a small fraction of source symbols remain uncovered by any output symbols, which means there exists a fraction of all zero columns in the coefficient matrix $\mathbf{M}$. These uncovered source symbols cannot be recovered at the end of the peeling decoding process. This implies that FLT codes do not perform well for moderate values of $N$. This is a well known issue with LT codes~\cite{LTcodes}. To address this issue, we propose a Raptor code based scheme, termed \emph{factored Raptor (FR) codes}, which are described below.

\subsubsection{Encoding}
\label{sec:FRcodes_encoding}
An FR code is an FLT code concatenated with an outer code. In the case of single user channel, the source symbols are encoded using the outer code, and the input symbols of the LT code are formed by a codeword of the outer code. Unlike the case of single user erasure channel, the outer code of FR codes cannot be any arbitrary code, due to the structure imposed by the matrix-matrix multiplication problem. In distributed matrix multiplication problem, the worker nodes are asked to compute product of a linear combination of chunks of one matrix with other. Collection of these products should form a codeword of the FLT code, and the input symbols corresponding to the codeword of the FLT code should form a codeword of the outer code. This requirement is met by encoding $\mathbf{A}$ and $\mathbf{B}$ with an MDS code and sending the linear combination of chunks of the respective encoded matrix to the worker nodes.  
We briefly describe the encoding process below.
\begin{itemize}
    \item [1.] Encode the input matrix $\mathbf{A}=[\mathbf{A}_1,\mathbf{A}_2, \cdots,\mathbf{A}_m]$ using an $(\tilde{m},m)$ MDS code to obtain $\mathbf{\widetilde{A}}=[\mathbf{A}_1, \mathbf{A}_2, \cdots \mathbf{A}_{\tilde{m}}]$. Similarly, encode $\mathbf{B}$ to $\mathbf{\widetilde{B}}$
    using an $(\tilde{n},n)$ MDS code.
    \item [2.] Encode $\widetilde{\mathbf{A}}$ and $\widetilde{\mathbf{B}}$ according to the degree distribution of an FLT code as described in Section~\ref{sec:LTcodes}. 
\end{itemize}
Note that $\mathbf{\widetilde{C}}$ is a codeword of the FLT code, and  $\mathbf{A}_i\mathbf{B}_j$'s for $i \in [\tilde{m}]$ and $j \in [\tilde{n}]$ form a codeword of an $(\tilde{m},m)\times (\tilde{n},n)$ outer Product code.
\subsubsection{Decoding}
\label{FRcodes_decoding}
Decoding of FR codes consists of alternating decoding iterations on the FLT code and the outer Product code.
At the master node, encoding of FR codes induces a Tanner graph which has $\mathbf{A}_i\mathbf{B}_j$'s for $i \in [\tilde{m}]$ and $j \in [\tilde{n}]$ as the input symbols and $\mathbf{\widetilde{C}}^p$'s as the output symbols.
Decoding iterations of the FLT proceed on this Tanner graph as described in Section~\ref{sec:LTcodes_decoding}. 
The matrix $\mathbf{U}=\mathbf{\widetilde{A}}^\mathsf{T}\mathbf{\widetilde{B}}$ is a codeword of the $(\tilde{m},m)\times (\tilde{n},n)$ outer Product code. Each row and column of $\mathbf{U}$ is a codeword of the $(\tilde{n},n)$ MDS code and $(\tilde{m},m)$ MDS code, respectively. Decoding of the outer Product code proceeds by decoding the component MDS codes. 
Decoding of FR codes is described in Algorithm~1.
\begin{algorithm}[h]
\DontPrintSemicolon
\KwResult{Recover $\mathbf{C}$ from $\{\mathbf{\widetilde{C}}^p:p\in [N]\}$}
Find a row $\mathbf{M}_{p',:}$ in the matrix $\mathbf{M}$ with $\norm{\mathbf{M}_{p',:}}_0=1$.\;
Recover $\mathbf{A}^\mathsf{T}_{i}\mathbf{B}_j$ from $\mathbf{\widetilde{C}}^{p'}$.\;
Update $\mathbf{\widetilde{\underbar C}}=\mathbf{\widetilde{\underbar C}}-\mathbf{M}_{:,k}\mathbf{A}^\mathsf{T}_{i}\mathbf{B}_j$, where $k=(i-1)m+j$. Then, set $\mathbf{M}_{:,k}=\underbar 0$, where $\underbar 0$ is an all-zero vector of length $N$.\;
Repeat Steps $1$ to $3$ if there exists a row $\mathbf{M}_{p',:}$ in the matrix $\mathbf{M}$ such that $\norm{\mathbf{M}_{p',:}}_0=1$. Otherwise, go to the next step.\;  
Construct the matrix $\mathbf{U}=\mathbf{\widetilde{A}^\mathsf{T}\widetilde{B}}$ as follows: For $i \in [\tilde{m}]$ and $j \in [\tilde{n}] $, set $\mathbf{U}_{ij}=\mathbf{A}^T_i\mathbf{B}_j$ if the source symbol $\mathbf{A}^T_i\mathbf{B}_j$ is recovered. Otherwise, set $\mathbf{U}_{ij}$ as an erasure.\;
If  $\mathbf{U}_{:,j}$, for $j \in [\tilde{n}]$, has less than $\tilde{m}-m+1$ erasures, decode it using the $(\tilde{m},m)$ MDS code.\;
Similarly, if  $\mathbf{U}_{i,:}$, for $i \in [\tilde{m}]$, has less than $\tilde{n}-n+1$ erasures, decode it using the $(\tilde{n},n)$ MDS code.\;
For all recovered symbols $\mathbf{A}_i\mathbf{B}_j$  in Steps $6$ and $7$, update $\mathbf{\widetilde{\underbar C}}=\mathbf{\widetilde{\underbar C}}-\mathbf{M}_{:,k}$, where $k=(i-1)m+j$. Then, set $\mathbf{M}_{:,k}=\underbar 0$. \;
Repeat Steps $6, 7$ and $8$ until there exists a row (column) with less than $\tilde{n}-n+1$ ($\tilde{m}-m+1$) 
erasures.\; 
 If the number of unrecovered symbols at the end of Step $8$ is less than that of Step $4$, go to Step $1$. Otherwise, go to the next step.\;
 
 If $\mathbf{A}_i^\mathsf{T}\mathbf{B}_j$ for all $i \in [m]$ and $j \in [n]$ are recovered, declare the decoding process as success. Otherwise, declare a decoding failure.
 
 \caption{Decoding process at the master node}
 \label{alg:algorithm1}
\end{algorithm}
\begin{example}
\label{ex:example1}
 We now illustrate encoding of an FR code for multiplication of $\mathbf{A}=[\mathbf{A}_1, \mathbf{A}_2]$ and $\mathbf{B}=[\mathbf{B}_1, \mathbf{B}_2]$. 
For this purpose, We consider an $(3,2)^2$ Product code as the outer code with a $(3,2)$ MDS code as the component code, and a $(10,9)$ FLT code with degree distribution $\Omega(x)=0.2x+0.7x^2+0.1x^4$. Encoding of the outer Product code is done by applying the $(3,2)$ MDS code to both $\mathbf{A}$ and $\mathbf{B}$ to obtain $\mathbf{\widetilde{A}}=[\mathbf{{A}}_1, \mathbf{{A}}_2, \mathbf{{A}}_3]$ and $\mathbf{\widetilde{B}}=[\mathbf{{B}}_1, \mathbf{{B}}_2, \mathbf{{B}}_3]$, respectively. Recall that encoding of a degree-$4$ output symbol of a FLT code is shown in Example~\ref{ex:Ltencoding}.
Similarly, encoding of the FLT code is done by asking worker node $p$ for $p \in [10]$ to compute $\mathbf{\widetilde{C}}^p$ according to $\Omega(x)$ as follows:
\begin{align*}
   & \widetilde{\mathbf{C}}^1=\mathbf{A}_1^{\mathsf{T}}\mathbf{B}_2, \quad \widetilde{\mathbf{C}}^2=\mathbf{A}_1^{\mathsf{T}}\mathbf{B}_3,\\
   &  \widetilde{\mathbf{C}}^3= (\mathbf{A}_1^{\mathsf{T}}+\mathbf{A}_2^{\mathsf{T}})\mathbf{B}_1,  \quad \widetilde{\mathbf{C}}^4=  (\mathbf{A}_1^{\mathsf{T}}+\mathbf{A}_3^{\mathsf{T}})\mathbf{B}_1,\\
   &  \widetilde{\mathbf{C}}^5= (\mathbf{A}_1^{\mathsf{T}}+\mathbf{A}_2^{\mathsf{T}})\mathbf{B}_3,  \quad \widetilde{\mathbf{C}}^6= (\mathbf{A}_1^{\mathsf{T}}+\mathbf{A}_3^{\mathsf{T}})\mathbf{B}_3,\\
  &  \widetilde{\mathbf{C}}^7= \mathbf{A}_1^{\mathsf{T}}(\mathbf{B}_1+\mathbf{B}_2),  \quad \widetilde{\mathbf{C}}^8= \mathbf{A}_2^{\mathsf{T}}(\mathbf{B}_2+\mathbf{B}_3),\\
   & \widetilde{\mathbf{C}}^9= \mathbf{A}_3^{\mathsf{T}}(\mathbf{B}_1+\mathbf{B}_2),  \quad \widetilde{\mathbf{C}}^{10}= (\mathbf{A}_2^{\mathsf{T}}+\mathbf{A}_3^{\mathsf{T}})(\mathbf{B}_2+\mathbf{B}_3).
\end{align*}
Encoding of the outer Product code is shown in Fig.~\ref{fig:Encoding_prod}. The equivalent Tanner graph of the FLT is shown in Fig.~\ref{fig:Encoding_FLT}.
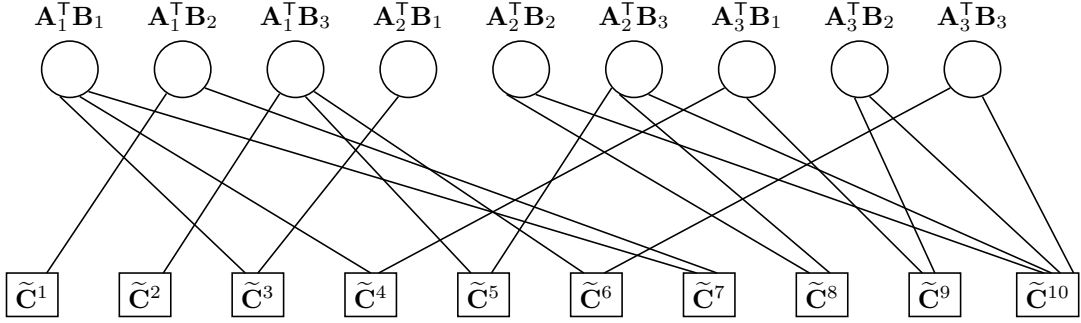
\begin{figure}[h]
    \centering
    \begin{subfigure}{0.23\textwidth}
  \begin{tikzpicture}[every node/.style={yscale=1.2}]
\begin{scope}[node distance=1.2cm,>=angle 90,semithick]
\node[conode] (v1) {$\mathbf{A}_1^{\mathsf{T}}\mathbf{B}_1$};
\node[conode] (v2)[right of=v1] {$\mathbf{A}_1^{\mathsf{T}}\mathbf{B}_2$};
\node[rnode] (v3)[right of=v2] {$\mathbf{A}_1^{\mathsf{T}}\mathbf{B}_3$};
\node[conode] (v4)[below of=v1,yshift=0.5cm] {$\mathbf{A}_2^{\mathsf{T}}\mathbf{B}_1$};
\node[conode] (v5)[below of=v2,yshift=0.5cm] {$\mathbf{A}_2^{\mathsf{T}}\mathbf{B}_2$};
\node[rnode] (v6)[below of=v3,yshift=0.5cm] {$\mathbf{A}_2^{\mathsf{T}}\mathbf{B}_3$};
\node[rnode] (v7)[below of=v4,yshift=0.5cm] {$\mathbf{A}_3^{\mathsf{T}}\mathbf{B}_1$};
\node[rnode] (v8)[below of=v5,yshift=0.5cm] {$\mathbf{A}_3^{\mathsf{T}}\mathbf{B}_2$};
\node[rnode] (v9)[below of=v6,yshift=0.5cm] {$\mathbf{A}_3^{\mathsf{T}}\mathbf{B}_3$};
\end{scope}    
\end{tikzpicture}
  \caption{$(3,2)^2$ outer Product code}
  \label{fig:Encoding_prod}
    \end{subfigure}\vspace{0.5cm}
    \begin{subfigure}{0.82\textwidth}
    \begin{tikzpicture}[every node/.style={scale=1}]
\begin{scope}[node distance=2 cm,>=angle 90,semithick]
\node[cnode] (v1) {};
\node[] (l1) [above of=v1,yshift=-1.3cm,] {$\mathbf{A}_1^{\mathsf{T}}\mathbf{B}_1$};
\node[cnode] (v2)[right of=v1,xshift=-0.5cm] {};
\node[] (l2) [above of=v2,yshift=-1.3cm] {$\mathbf{A}_1^{\mathsf{T}}\mathbf{B}_2$};
\node[cnode] (v3)[right of=v2, xshift=-0.5cm] {};
\node[] (l3) [above of=v3,yshift=-1.3cm] {$\mathbf{A}_1^{\mathsf{T}}\mathbf{B}_3$};
\node[cnode] (v4)[right of=v3,yshift=0cm,xshift=-0.5cm] {};
\node[] (l4) [above of=v4,yshift=-1.3cm] {$\mathbf{A}_2^{\mathsf{T}}\mathbf{B}_1$};
\node[cnode] (v5)[right of=v4,yshift=0cm,xshift=-0.5cm] {};
\node[] (l5) [above of=v5,yshift=-1.3cm] {$\mathbf{A}_2^{\mathsf{T}}\mathbf{B}_2$};
\node[cnode] (v6)[right of=v5,yshift=0cm,xshift=-0.5cm] {};
\node[] (l6) [above of=v6,yshift=-1.3cm] {$\mathbf{A}_2^{\mathsf{T}}\mathbf{B}_3$};
\node[cnode] (v7)[right of=v6,yshift=0cm,xshift=-0.5cm] {};
\node[] (l7) [above of=v7,yshift=-1.3cm] {$\mathbf{A}_3^{\mathsf{T}}\mathbf{B}_1$};
\node[cnode] (v8)[right of=v7,yshift=0cm,xshift=-0.5cm] {};
\node[] (l8) [above of=v8,yshift=-1.3cm] {$\mathbf{A}_3^{\mathsf{T}}\mathbf{B}_2$};
\node[cnode] (v9)[right of=v8,yshift=0cm,xshift=-0.5cm] {};
\node[] (l9) [above of=v9,yshift=-1.3cm] {$\mathbf{A}_3^{\mathsf{T}}\mathbf{B}_3$};
\node[rnode] (c1) [below of=v1,xshift=-0.5cm,yshift=-1cm] {$\mathbf{\widetilde{C}}^1$};
\node[rnode] (c2) [below of=v2,xshift=-0.5cm,yshift=-1cm] {$\mathbf{\widetilde{C}}^2$};
\node[rnode] (c3) [below of=v3,xshift=-0.5cm,yshift=-1cm] {$\mathbf{\widetilde{C}}^3$};
\node[rnode] (c4) [below of=v4,xshift=-0.5cm,yshift=-1cm] {$\mathbf{\widetilde{C}}^4$};
\node[rnode] (c5) [below of=v5,xshift=-0.5cm,yshift=-1cm] {$\mathbf{\widetilde{C}}^5$};
\node[rnode] (c6) [below of=v6,xshift=-0.5cm,yshift=-1cm] {$\mathbf{\widetilde{C}}^6$};
\node[rnode] (c7) [below of=v7,xshift=-0.5cm,yshift=-1cm] {$\mathbf{\widetilde{C}}^7$};
\node[rnode] (c8) [below of=v8,xshift=-0.5cm,yshift=-1cm] {$\mathbf{\widetilde{C}}^8$};
\node[rnode] (c9) [below of=v9,xshift=-0.5cm,yshift=-1cm] {$\mathbf{\widetilde{C}}^9$};
\node[rnode] (c10) [below of=v9,xshift=1cm,yshift=-1cm] {$\mathbf{\widetilde{C}}^{10}$};
\draw[black] (c1) -- (v2);
\draw[black] (c2.50) -- (v3);
\draw[black] (c3.120) -- (v1.250);
\draw[black] (c3.90) -- (v4.250);
\draw[black] (c4.90) -- (v1.290);
\draw[black] (c4.70) -- (v7.220);
\draw[black] (c5.120) -- (v3.290);
\draw[black] (c5.70) -- (v6.220);
\draw[black] (c6.120) -- (v3.310);
\draw[black] (c6.70) -- (v9.220);
\draw[black] (c7.120) -- (v1.310);
\draw[black] (c7.70) -- (v2.320);
\draw[black] (c8.120) -- (v5.240);
\draw[black] (c8.70) -- (v6.240);
\draw[black] (c9.120) -- (v7.270);
\draw[black] (c9.90) -- (v8.260);
\draw[black] (c10.100) -- (v5.300);
\draw[black] (c10.80) -- (v6.300);
\draw[black] (c10.60) -- (v8.290);
\draw[black] (c10.40) -- (v9.290);
\end{scope}    
\end{tikzpicture}
    
    \caption{Tanner graph of the FLT code in FR code}
    \label{fig:Encoding_FLT}
    \end{subfigure}\vspace{0.25cm}
    \caption{Illustration of the encoding scheme for FR codes}
    \label{fig:Encoding}
\end{figure}

\end{example}

\begin{figure}[htb]
    \centering
    \begin{subfigure}{0.22\textwidth}
\begin{tikzpicture}[every node/.style={scale=0.7}]
\begin{scope}[node distance=2 cm,>=angle 90,semithick]
\node[cnode] (v1) {};
\node[] (l1) [left of=v1,xshift=1cm] {$\mathbf{A}_1^{\mathsf{T}}\mathbf{B}_1$};
\node[cnode] (v2) [below of=v1,yshift=1cm] {};
\node[] (l2) [left of=v2,xshift=1cm] {$\mathbf{A}_1^{\mathsf{T}}\mathbf{B}_2$};
\node[cnode] (v3) [below of=v2,yshift=1cm] {};
\node[] (l3) [left of=v3,xshift=1cm] {$\mathbf{A}_1^{\mathsf{T}}\mathbf{B}_3$};
\node[cnode] (v4) [below of=v3,yshift=1cm] {};
\node[] (l4) [left of=v4,xshift=1cm] {$\mathbf{A}_2^{\mathsf{T}}\mathbf{B}_1$};
\node[cnode] (v5) [below of=v4,yshift=1cm] {};
\node[] (l5) [left of=v5,xshift=1cm] {$\mathbf{A}_2^{\mathsf{T}}\mathbf{B}_3$};
\node[rnode] (c1) [right of=v1,yshift=-0.5cm,xshift=0.5cm] {$\mathbf{\widehat{C}}^1$};
\node[rnode] (c2) [below of=c1,yshift=1cm,xshift=0cm] {$\mathbf{\widetilde{C}}^3$};
\node[rnode] (c3) [below of=c2,yshift=1cm,xshift=0cm] {$\mathbf{\widetilde{C}}^5$};
\node[rnode] (c4) [below of=c3,yshift=1cm,xshift=0cm] {$\mathbf{\widetilde{C}}^7$};
\draw[black] (v1.0) -- (c2.160);
\draw[black] (v1.-30) -- (c3.160);
\draw[black] (v1.-50) -- (c4.180);
\draw[red] (v2) -- (c1);
\draw[black] (v2.-10) -- (c4.210);
\draw[black] (v3) -- (c3);
\draw[black] (v4) -- (c2.180);
\draw[black] (v5) -- (c3.200);
\end{scope}    
\end{tikzpicture}
\caption{Iteration $0$}
\label{fig:iteration0}
    \end{subfigure}
    \begin{subfigure}{0.22\textwidth}
   \begin{tikzpicture}[every node/.style={scale=0.7}]
\begin{scope}[node distance=2 cm,>=angle 90,semithick]
\node[cnode] (v1) {};
\node[] (l1) [left of=v1,xshift=1cm] {$\mathbf{A}_1^{\mathsf{T}}\mathbf{B}_1$};
\node[cnode] (v2) [below of=v1,yshift=1cm] {};
\node[] (l2) [left of=v2,xshift=1cm] {$\mathbf{A}_1^{\mathsf{T}}\mathbf{B}_2$};
\node[cnode] (v3) [below of=v2,yshift=1cm] {};
\node[] (l3) [left of=v3,xshift=1cm] {$\mathbf{A}_1^{\mathsf{T}}\mathbf{B}_3$};
\node[cnode] (v4) [below of=v3,yshift=1cm] {};
\node[] (l4) [left of=v4,xshift=1cm] {$\mathbf{A}_2^{\mathsf{T}}\mathbf{B}_1$};
\node[cnode] (v5) [below of=v4,yshift=1cm] {};
\node[] (l5) [left of=v5,xshift=1cm] {$\mathbf{A}_2^{\mathsf{T}}\mathbf{B}_3$};
\node[rnode] (c1) [right of=v1,yshift=-0.5cm,xshift=0.5cm] {$\mathbf{\widetilde{C}}^1$};
\node[rnode] (c2) [below of=c1,yshift=1cm,xshift=0cm] {$\mathbf{\widetilde{C}}^3$};
\node[rnode] (c3) [below of=c2,yshift=1cm,xshift=0cm] {$\mathbf{\widetilde{C}}^5$};
\node[rnode] (c4) [below of=c3,yshift=1cm,xshift=0cm] {$\mathbf{\widetilde{C}}^7$};
\draw[black] (v1.0) -- (c2.160);
\draw[black] (v1.-30) -- (c3.160);
\draw[red] (v1.-50) -- (c4.180);
\draw[black] (v3) -- (c3);
\draw[black] (v4) -- (c2.180);
\draw[black] (v5) -- (c3.200);
\end{scope}    
\end{tikzpicture}
\caption{Iteration $1$}
\label{fig:iteration1}
    \end{subfigure}
    \begin{subfigure}{0.22\textwidth}
   \begin{tikzpicture}[every node/.style={scale=0.7}]
\begin{scope}[node distance=2 cm,>=angle 90,semithick]
\node[cnode] (v1) {};
\node[] (l1) [left of=v1,xshift=1cm] {$\mathbf{A}_1^{\mathsf{T}}\mathbf{B}_1$};
\node[cnode] (v2) [below of=v1,yshift=1cm] {};
\node[] (l2) [left of=v2,xshift=1cm] {$\mathbf{A}_1^{\mathsf{T}}\mathbf{B}_2$};
\node[cnode] (v3) [below of=v2,yshift=1cm] {};
\node[] (l3) [left of=v3,xshift=1cm] {$\mathbf{A}_1^{\mathsf{T}}\mathbf{B}_3$};
\node[cnode] (v4) [below of=v3,yshift=1cm] {};
\node[] (l4) [left of=v4,xshift=1cm] {$\mathbf{A}_2^{\mathsf{T}}\mathbf{B}_1$};
\node[cnode] (v5) [below of=v4,yshift=1cm] {};
\node[] (l5) [left of=v5,xshift=1cm] {$\mathbf{A}_2^{\mathsf{T}}\mathbf{B}_3$};
\node[rnode] (c1) [right of=v1,yshift=-0.5cm,xshift=0.5cm] {$\mathbf{\widetilde{C}}^1$};
\node[rnode] (c2) [below of=c1,yshift=1cm,xshift=0cm] {$\mathbf{\widetilde{C}}^3$};
\node[rnode] (c3) [below of=c2,yshift=1cm,xshift=0cm] {$\mathbf{\widetilde{C}}^5$};
\node[rnode] (c4) [below of=c3,yshift=1cm,xshift=0cm] {$\mathbf{\widetilde{C}}^7$};
\draw[black] (v3) -- (c3);
\draw[red] (v4) -- (c2.180);
\draw[black] (v5) -- (c3.200);
\end{scope}    
\end{tikzpicture}
\caption{Iteration $2$}
\label{fig:iteration2}
    \end{subfigure}
    \begin{subfigure}{0.22\textwidth}
   \begin{tikzpicture}[every node/.style={scale=0.7}]
\begin{scope}[node distance=2 cm,>=angle 90,semithick]
\node[cnode] (v1) {};
\node[] (l1) [left of=v1,xshift=1cm] {$\mathbf{A}_1^{\mathsf{T}}\mathbf{B}_1$};
\node[cnode] (v2) [below of=v1,yshift=1cm] {};
\node[] (l2) [left of=v2,xshift=1cm] {$\mathbf{A}_1^{\mathsf{T}}\mathbf{B}_2$};
\node[cnode] (v3) [below of=v2,yshift=1cm] {};
\node[] (l3) [left of=v3,xshift=1cm] {$\mathbf{A}_1^{\mathsf{T}}\mathbf{B}_3$};
\node[cnode] (v4) [below of=v3,yshift=1cm] {};
\node[] (l4) [left of=v4,xshift=1cm] {$\mathbf{A}_2^{\mathsf{T}}\mathbf{B}_1$};
\node[cnode] (v5) [below of=v4,yshift=1cm] {};
\node[] (l5) [left of=v5,xshift=1cm] {$\mathbf{A}_2^{\mathsf{T}}\mathbf{B}_3$};
\node[rnode] (c1) [right of=v1,yshift=-0.5cm,xshift=0.5cm] {$\mathbf{\widetilde{C}}^1$};
\node[rnode] (c2) [below of=c1,yshift=1cm,xshift=0cm] {$\mathbf{\widetilde{C}}^3$};
\node[rnode] (c3) [below of=c2,yshift=1cm,xshift=0cm] {$\mathbf{\widetilde{C}}^5$};
\node[rnode] (c4) [below of=c3,yshift=1cm,xshift=0cm] {$\mathbf{\widetilde{C}}^7$};
\draw[black] (v3) -- (c3);
\draw[black] (v5) -- (c3.200);
\end{scope}    
\end{tikzpicture}
\caption{Iteration $3$}
\label{fig:iteration3}
    \end{subfigure}\vspace{0.75cm}
    \begin{subfigure}{0.24\textwidth}
    \begin{tikzpicture}[every node/.style={yscale=1.4}]
\begin{scope}[node distance=1.2cm,>=angle 90,semithick]
\node[prnode] (v1) {$\mathbf{A}_1^{\mathsf{T}}\mathbf{B}_1$};
\node[prnode] (v2)[right of=v1] {$\mathbf{A}_1^{\mathsf{T}}\mathbf{B}_2$};
\node[fnode] (v3)[right of=v2] {\large \bf{?}};
\node[prnode] (v4)[below of=v1,yshift=0.5cm] {$\mathbf{A}_2^{\mathsf{T}}\mathbf{B}_1$};
\node[prnode] (v5)[below of=v2,yshift=0.5cm] {\large \bf{?}};
\node[prnode] (v6)[below of=v3,yshift=0.5cm] {\large \bf{?}};
\node[fnode] (v7)[below of=v4,yshift=0.5cm] {\large \bf{?}};
\node[prnode] (v8)[below of=v5,yshift=0.5cm] {\large \bf{?}};
\node[prnode] (v9)[below of=v6,yshift=0.5cm] {\large \bf{?}};
\end{scope}    
\end{tikzpicture}
\caption{Iteration $4$}
\label{fig:iteration4}
    \end{subfigure}
    \begin{subfigure}{0.24\textwidth}
    \begin{tikzpicture}[every node/.style={yscale=1.4}]
\begin{scope}[node distance=1.2cm,>=angle 90,semithick]
\node[prnode] (v1) {$\mathbf{A}_1^{\mathsf{T}}\mathbf{B}_1$};
\node[prnode] (v2)[right of=v1] {$\mathbf{A}_1^{\mathsf{T}}\mathbf{B}_2$};
\node[prnode] (v3)[right of=v2] {$\mathbf{A}_1^{\mathsf{T}}\mathbf{B}_3$};
\node[prnode] (v4)[below of=v1,yshift=0.5cm] {$\mathbf{A}_2^{\mathsf{T}}\mathbf{B}_1$};
\node[prnode] (v5)[below of=v2,yshift=0.5cm] {\large \bf{?}};
\node[prnode] (v6)[below of=v3,yshift=0.5cm] {\large \bf{?}};
\node[prnode] (v7)[below of=v4,yshift=0.5cm] {$\mathbf{A}_3^{\mathsf{T}}\mathbf{B}_1$};
\node[prnode] (v8)[below of=v5,yshift=0.5cm] {\large \bf{?}};
\node[prnode] (v9)[below of=v6,yshift=0.5cm] {\large \bf{?}};
\end{scope}    
\end{tikzpicture}
    \caption{Iteration $5$}
    \label{fig:iteration5}
    \end{subfigure}\vspace{0.75cm}
     \begin{subfigure}{0.22\textwidth}
   \begin{tikzpicture}[every node/.style={scale=0.7}]
\begin{scope}[node distance=2 cm,>=angle 90,semithick]
\node[cnode] (v1) {};
\node[] (l1) [left of=v1,xshift=1cm] {$\mathbf{A}_1^{\mathsf{T}}\mathbf{B}_1$};
\node[cnode] (v2) [below of=v1,yshift=1cm] {};
\node[] (l2) [left of=v2,xshift=1cm] {$\mathbf{A}_1^{\mathsf{T}}\mathbf{B}_2$};
\node[cnode] (v3) [below of=v2,yshift=1cm] {};
\node[] (l3) [left of=v3,xshift=1cm] {$\mathbf{A}_1^{\mathsf{T}}\mathbf{B}_3$};
\node[cnode] (v4) [below of=v3,yshift=1cm] {};
\node[] (l4) [left of=v4,xshift=1cm] {$\mathbf{A}_2^{\mathsf{T}}\mathbf{B}_1$};
\node[cnode] (v5) [below of=v4,yshift=1cm] {};
\node[] (l5) [left of=v5,xshift=1cm] {$\mathbf{A}_2^{\mathsf{T}}\mathbf{B}_3$};
\node[rnode] (c1) [right of=v1,yshift=-0.5cm,xshift=0.5cm] {$\mathbf{\widetilde{C}}^1$};
\node[rnode] (c2) [below of=c1,yshift=1cm,xshift=0cm] {$\mathbf{\widetilde{C}}^3$};
\node[rnode] (c3) [below of=c2,yshift=1cm,xshift=0cm] {$\mathbf{\widetilde{C}}^5$};
\node[rnode] (c4) [below of=c3,yshift=1cm,xshift=0cm] {$\mathbf{\widetilde{C}}^7$};
\draw[red] (v5) -- (c3.200);
\end{scope}    
\end{tikzpicture}
\caption{Iteration $6$}
\label{fig:iteration6}
    \end{subfigure}
    \begin{subfigure}{0.24\textwidth}
    \begin{tikzpicture}[every node/.style={yscale=1.4}]
\begin{scope}[node distance=1.2cm,>=angle 90,semithick]
\node[prnode] (v1) {$\mathbf{A}_1^{\mathsf{T}}\mathbf{B}_1$};
\node[prnode] (v2)[right of=v1] {$\mathbf{A}_1^{\mathsf{T}}\mathbf{B}_2$};
\node[prnode] (v3)[right of=v2] {$\mathbf{A}_1^{\mathsf{T}}\mathbf{B}_3$};
\node[prnode] (v4)[below of=v1,yshift=0.5cm] {$\mathbf{A}_2^{\mathsf{T}}\mathbf{B}_1$};
\node[fnode] (v5)[below of=v2,yshift=0.5cm] {\large \bf{?}};
\node[prnode] (v6)[below of=v3,yshift=0.5cm] {$\mathbf{A}_2^{\mathsf{T}}\mathbf{B}_3$};
\node[prnode] (v7)[below of=v4,yshift=0.5cm] {$\mathbf{A}_3^{\mathsf{T}}\mathbf{B}_1$};
\node[prnode] (v8)[below of=v5,yshift=0.5cm] {\large \bf{?}};
\node[prnode] (v9)[below of=v6,yshift=0.5cm] {\large \bf{?}};
\end{scope}    
\end{tikzpicture}
\caption{Iteration $7$}
\label{fig:iteration7}
    \end{subfigure}
    \begin{subfigure}{0.24\textwidth}
    \begin{tikzpicture}[every node/.style={yscale=1.4}]
\begin{scope}[node distance=1.2cm,>=angle 90,semithick]
\node[prnode] (v1) {$\mathbf{A}_1^{\mathsf{T}}\mathbf{B}_1$};
\node[prnode] (v2)[right of=v1] {$\mathbf{A}_1^{\mathsf{T}}\mathbf{B}_2$};
\node[prnode] (v3)[right of=v2] {$\mathbf{A}_1^{\mathsf{T}}\mathbf{B}_3$};
\node[prnode] (v4)[below of=v1,yshift=0.5cm] {$\mathbf{A}_2^{\mathsf{T}}\mathbf{B}_1$};
\node[prnode] (v5)[below of=v2,yshift=0.5cm] {$\mathbf{A}_2^{\mathsf{T}}\mathbf{B}_2$};
\node[prnode] (v6)[below of=v3,yshift=0.5cm] {$\mathbf{A}_2^{\mathsf{T}}\mathbf{B}_3$};
\node[prnode] (v7)[below of=v4,yshift=0.5cm] {$\mathbf{A}_3^{\mathsf{T}}\mathbf{B}_1$};
\node[fnode] (v8)[below of=v5,yshift=0.5cm] {\large \bf{?}};
\node[fnode] (v9)[below of=v6,yshift=0.5cm] {\large \bf{?}};
\end{scope}    
\end{tikzpicture}
\caption{Iteration $8$}
\label{fig:iteration8}
    \end{subfigure}
    \begin{subfigure}{0.24\textwidth}
    \begin{tikzpicture}[every node/.style={yscale=1.4}]
\begin{scope}[node distance=1.2cm,>=angle 90,semithick]
\node[prnode] (v1) {$\mathbf{A}_1^{\mathsf{T}}\mathbf{B}_1$};
\node[prnode] (v2)[right of=v1] {$\mathbf{A}_1^{\mathsf{T}}\mathbf{B}_2$};
\node[prnode] (v3)[right of=v2] {$\mathbf{A}_1^{\mathsf{T}}\mathbf{B}_3$};
\node[prnode] (v4)[below of=v1,yshift=0.5cm] {$\mathbf{A}_2^{\mathsf{T}}\mathbf{B}_1$};
\node[prnode] (v5)[below of=v2,yshift=0.5cm] {$\mathbf{A}_2^{\mathsf{T}}\mathbf{B}_2$};
\node[prnode] (v6)[below of=v3,yshift=0.5cm] {$\mathbf{A}_2^{\mathsf{T}}\mathbf{B}_3$};
\node[prnode] (v7)[below of=v4,yshift=0.5cm] {$\mathbf{A}_3^{\mathsf{T}}\mathbf{B}_1$};
\node[prnode] (v8)[below of=v5,yshift=0.5cm] {$\mathbf{A}_3^{\mathsf{T}}\mathbf{B}_2$};
\node[prnode] (v9)[below of=v6,yshift=0.5cm] {$\mathbf{A}_3^{\mathsf{T}}\mathbf{B}_3$};
\end{scope}    
\end{tikzpicture}
\caption{Iteration $9$}
\label{fig:iteration9}
    \end{subfigure}\vspace{0.25cm}
    \caption{Illustration of the decoding algorithm for FR codes}
    \label{fig:decoding}
\end{figure}
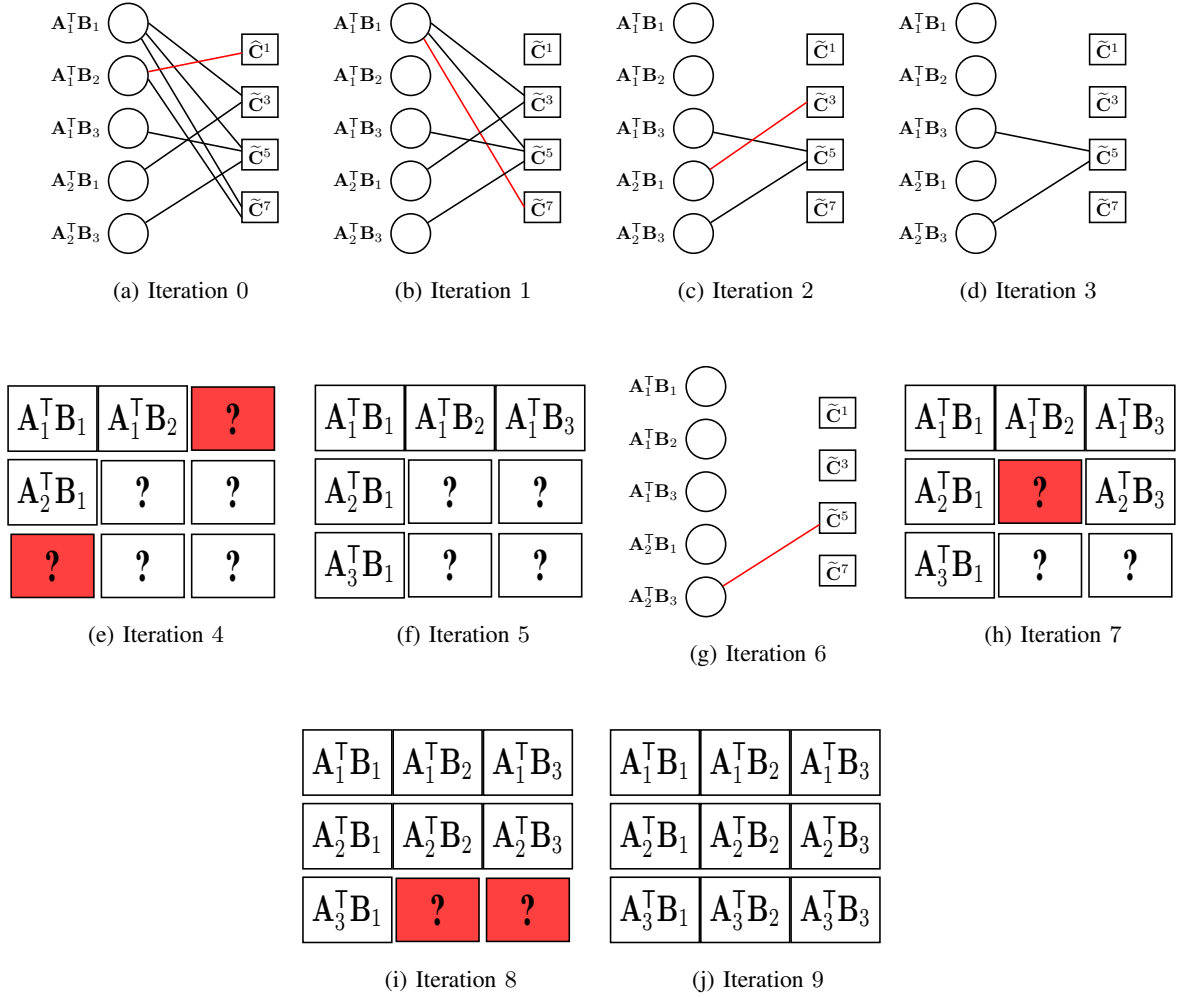

\begin{example}
\label{ex:example2}
In this example, we describe the decoding algorithm at the master node using the code described in Example~\ref{ex:example1}.
For this example, consider that the master node collects results from worker node $\{1,3,5,7\}$. The peeling decoding is run on the subgraph, denoted by $G_0$, induced by worker node $\{1,3,5,7\}.$ In iteration $0$,  the source symbol $\mathbf{A}_1^{\mathsf{T}}\mathbf{B}_2$ is recovered from  worker $1$ since $\mathbf{\widetilde{C}}^1$ is a degree-$1$ node in $G_0$. Peel the edges connected to the source symbol $\mathbf{A}_1^{\mathsf{T}}\mathbf{B}_2$ in $G_0$ and denote the residual graph by $G_1$. In iteration $1$, the source symbol $\mathbf{A}_1^{\mathsf{T}}\mathbf{B}_1=\mathbf{\widetilde{C}}^7-\mathbf{A}_1^{\mathsf{T}}\mathbf{B}_2$ is recovered from worker $1$ since $\mathbf{\widetilde{C}}^1$ is a degree-$1$ node in $G_1$. Similarly, the source symbol $\mathbf{A}_2^{\mathsf{T}}\mathbf{B}_1$ is recovered from  worker $3$ in iteration $3$. Since there are no degree-1 symbols left after iteration $3$, the decoder of the FLT code cannot proceed further. Decoding of the FR code can proceed further by decoding the outer Product code. Arrange the input symbols to the FLT code in a $3 \times 2$ matrix as shown in Fig.~\ref{fig:iteration4}. Both first row and first column have two computation results as shown in Fig.~\ref{fig:iteration4}. In iteration~$4$, the missing computation results in both first row and first column are recovered by decoding the $(3,2)$ MDS code.  Similarly, the decoding process continues as shown in Fig.'s~\ref{fig:iteration5}-\ref{fig:iteration9}.
\end{example}
\subsubsection{Optimal Decoding by Inactivation Decoding}
\label{sec:optimal_decoding}
Maximum likelihood decoding of FLT and FR codes can be implemented using a low-complexity algorithm known as inactivation decoding \cite{Burshtein}, which is a combination of peeling decoding and matrix inversion. In particular, the inactivation decoding starts with peeling decoding, referred to as the first phase of decoding, which continues until it encounters a stopping set. At this point, the decoder assumes that it knows the value of an unrecovered symbol, referred to as an \emph{inactivated symbol}. Then, the peeling decoder starts again and computes the unrecovered symbols in terms of the value of the inactivated symbol. This procedure is repeated again, when the peeling decoder is stuck, by choosing another inactivated symbol from the unrecovered input symbols. The decoding stops when all the input symbols are either recovered or inactivated.  Finally, optimal decoding of the inactivated symbols is performed via Gaussian elimination, and the decoded values are back-substituted into the decoded input symbols which depend on them. Let $\mathbf{R}$ denote the residual coefficient matrix of the FLT code after removing all zero columns and rows at the end of the first phase of decoding. Let $\mathbf{G}$ denote the generator matrix of the outer Product code with rows and columns corresponding to the recovered symbols removed. Let $ \mathbf{C}'$ denote the set of residual output symbols whose degree is at least two at the end of the first phase of decoding. Let $\mathbf{Y}$ denote the set of inactivated symbols. Without loss of generality, assume that the input symbols corresponding to the first $|\mathbf{Y}|$ columns of $\mathbf{R}$ are inactivated.  Let $\mathbf{Z}$ denote the set of input symbols recovered in terms of $\mathbf{Y}$. Then, the following equation holds:
\begin{equation}
\label{eq:4}
    \mathbf{R}\mathbf{G} \begin{bmatrix}
        \mathbf{Y} \\
        \mathbf{Z}\\
    \end{bmatrix}=\mathbf{C}'.
\end{equation}
At the end of decoding, $\mathbf{Z}$ can be expressed in terms of $\mathbf{Y}$ as follows:
\begin{equation}
\label{eq:5}
    \mathbf{Z}= \mathbf{D} \mathbf{Y} + \mathbf{X}.
\end{equation}
Combining~\eqref{eq:4} and~\eqref{eq:5}, we get
\begin{equation}
    \mathbf{Q}\mathbf{Y} + \mathbf{R}\mathbf{G} \mathbf{X}=\mathbf{C}'
\end{equation} where $\mathbf{Q} =\mathbf{R}\mathbf{G} \begin{bmatrix}
        \mathbf{I}\\
        \mathbf{D}  \\
    \end{bmatrix}$. If the matrix $\mathbf{Q}$ is full rank, the inactivated symbols can be recovered as follows:
 \begin{equation}
     \mathbf{Y}=\mathbf{Q}^{-1}\left(\mathbf{R}\mathbf{G} \mathbf{X}-\mathbf{C}'\right).
 \end{equation}
 Therefore, the decoding is successful if the matrix $\mathbf{Q}$ is full rank.
 
 \section{Computation and Communication Cost}
 In our encoding scheme, the master node sends two matrices $\widetilde{\mathbf{A}}^p$ and $\widetilde{\mathbf{B}}^p$ to each worker node $p$, and each worker node  computes only one product. 
 Therefore, the computation and communication costs of our proposed scheme is the same as that of
Polynomial codes and Product codes. 
Unlike our proposed FLT coding scheme, the LT coding scheme proposed in~\cite{Shroff} requires the master node to send (on average) $\log K$  chunks of the input matrices to each worker node, and each worker node computes (on average) $\log K$ products. 

\section{Decoding Complexity}
\label{sec:decoding_complexity}
In this section, we briefly describe the decoding complexity of FR codes. 
The peeling of every edge in the Tanner graph corresponds to performing $\frac{rt}{K}$ operations.
Let us assume that the average degree of output degree distribution is $d_{\text{avg}}$. Then each block $\mathbf{A}_i^\mathsf{T}\mathbf{B}_j$  will be involved in  $\mathcal{O}(\frac{d_{\text{avg}}N}{K}$) such operations on average. We are interested in the case when $\tilde{m}-m$ and $\tilde{n}-n$ are very small and hence, the complexity of decoding the outer code can be assumed to be small in comparison to that of decoding the LT part of the FR code. In addition, we are interested in a regime where $r$ and $t$ are very large and hence, 
the approximate complexity of the decoding algorithm is  $\mathcal{O}\left(rt\frac{d_{\text{avg}}N}{K}\right)$.

\section{Simulation Results}
In this section, we present simulation results to show that the recovery threshold of FR codes is higher than that of \emph{3-D} Product codes. To illustrate this, we choose $m=n=80$. We encode the source symbols $\mathbf{A}_i^{\mathsf{T}}$, for $i \in [m]$ and $j \in [n]$, using an $(82,80)^2$ Product code, where an $(82,80)$ MDS code is used as the row and column code. Then, the output symbols from the encoder of the Product code are encoded using an $(10000,6724)$ FLT code with right degree distribution $\Omega(x)=0.013x+0.5x^2+ 0.1661x^3+0.0726x^4+0.0826x^5+0.0581x^8+0.0340x^9+0.0576x^{18}+0.0160x^{66}$.
To illustrate the superiority of the encoding scheme in Section~\ref{sec:LTcodes}, we generate~coefficient matrix $\mathbf{M}$ in three different ways by altering the second step of encoding as follows: (Scheme I) Fix $d_1=d$ and $d_2=1$; (Scheme II) Define a uniform random variable $i$ which takes value from the set $\mathcal{I}=\{1,2\}$. Fix $d_i=d$ and $d_{\mathcal{I} \setminus \{i\}}=1$; (Scheme III) Follow the encoding process in Section~\ref{sec:LTcodes}. 
\begin{figure}[t]
    \centering
    \begin{tikzpicture}
\definecolor{mycolor1}{rgb}{0.63529,0.07843,0.18431}%
\definecolor{mycolor2}{rgb}{0.00000,0.44706,0.74118}%
\definecolor{mycolor3}{rgb}{0.00000,0.49804,0.00000}%
\definecolor{mycolor4}{rgb}{0.87059,0.49020,0.00000}%
\definecolor{mycolor5}{rgb}{0.00000,0.44700,0.74100}%
\definecolor{mycolor6}{rgb}{0.74902,0.00000,0.74902}%

\begin{semilogyaxis}[
xmin=2600, xmax=3150, ymin=1e-5, ymax=1, 
xlabel= Number of straggling nodes, ylabel= Probability of decoding failure,legend pos= outer north east,legend cell align=left,legend style={nodes={scale=0.75, transform shape}}]

\addplot [color=mycolor2,solid,line width=2.0pt,mark size=1.4pt,mark=square,mark options={solid}]
  table[row sep=crcr]{
2900 0.39\\
2850 0.03\\
2780 0.0008\\
2725 3e-5\\
};
\addlegendentry{\emph{3-D} Product code};

\addplot [color=mycolor4,solid,line width=2.0pt,mark size=1.3pt,mark=triangle,mark options={solid,rotate=90}]
  table[row sep=crcr]{
  2850 0.5\\
  2800 0.07\\
  2750 0.004\\
  2700 2e-4\\
};
\addlegendentry{FR code (Scheme I)};

\addplot [color=mycolor1,solid,line width=2.0pt,mark size=1.6pt,mark=diamond,mark options={solid}]
  table[row sep=crcr]{
 2900 0.3\\
 2850 0.02\\
 2800 0.0005\\
 2750 2.3e-5\\
};
\addlegendentry{FR code (Scheme II)};

\addplot [color=mycolor3,dashed,line width=2.0pt,mark size=1.4pt,mark=square,mark options={solid}]
  table[row sep=crcr]{
  3100 0.1\\
  3050 0.01\\
  2980 4e-4\\
  2940 4e-5\\
};
\addlegendentry{FR code (Scheme III)};

\end{semilogyaxis}

\end{tikzpicture}%
    \caption{Probability of decoding failure versus number of straggling nodes}
    \label{fig:LT_simulation_results}\vspace{-0.25cm}
\end{figure}
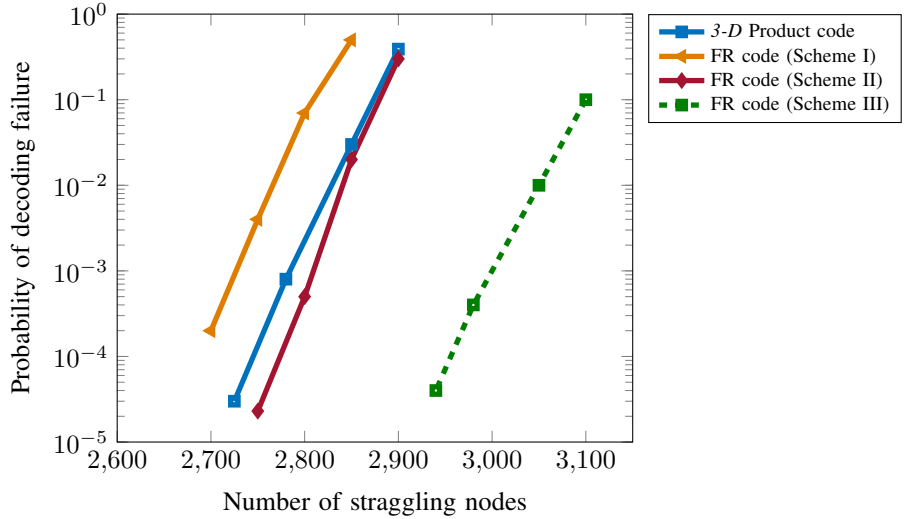 For each of these encoding schemes, in Fig.~\ref{fig:LT_simulation_results}, we have plotted the probability that the master node is unable to recover the output matrix $\mathbf{C}$ after receiving computations from $N=P-S$ workers, where $P$ is the total number of worker nodes and $S$ is the number of straggling worker nodes. We have also plotted the decoding failure of an $(21,18)\times(22,19) \times (22,19)$ Product code in Fig.~\ref{fig:LT_simulation_results} for comparison. We observe that the encoding scheme described in Section~\ref{sec:LTcodes} (Scheme III) has better recovery threshold when compared to the other two encoding schemes (Schemes I and II). In addition, one can observe that FR codes (Scheme III) have recovery threshold higher than \emph{3-D} Product codes.

\begin{figure}[t]
    \centering
    \begin{tikzpicture}
\definecolor{mycolor1}{rgb}{0.63529,0.07843,0.18431}%
\definecolor{mycolor2}{rgb}{0.00000,0.44706,0.74118}%
\definecolor{mycolor3}{rgb}{0.00000,0.49804,0.00000}%
\definecolor{mycolor4}{rgb}{0.87059,0.49020,0.00000}%
\definecolor{mycolor5}{rgb}{0.00000,0.44700,0.74100}%
\definecolor{mycolor6}{rgb}{0.74902,0.00000,0.74902}%
\pgfplotsset{set layers}
\begin{axis}[
axis y line*=left,
xmin=3050, xmax=3250, ymin=0, ymax=22, 
xlabel= Number of straggling nodes, ylabel= Average number of inactivated symbols,legend pos=north west,
legend style={nodes={scale=0.75, transform shape}}]
\addplot [color=mycolor1,solid,line width=2.0pt,mark size=1.4pt,mark=square,mark options={solid}]
  table[row sep=crcr]{
3250 22\\
3200 8.64\\
3150 2.47\\
3050 0.04\\
};\label{plot_one}
\end{axis}
\begin{semilogyaxis}[
xmin=3100,xmax=3250,ymax=8e-2,ymin=2e-4,
axis y line*=right,
axis x line=none,
ylabel style = {align=center},
ylabel={Probability of decoding failure},
legend pos=north west
]
\addlegendimage{/pgfplots/refstyle=plot_one}\addlegendentry{Number of inactivated symbols};
\addplot [color=mycolor3,solid,line width=1.0pt,mark size=1.4pt,mark=square,mark options={solid}]
  table[row sep=crcr]{
3250 0.02\\
3200 7e-3\\
3150 4e-4\\
};
\addlegendentry{Error rate};
\end{semilogyaxis}
\end{tikzpicture}%
    \caption{Average number of inactivated symbols versus number of straggling nodes}
    \label{fig:LT_optcode_complexity}\vspace{-0.25cm}
\end{figure}
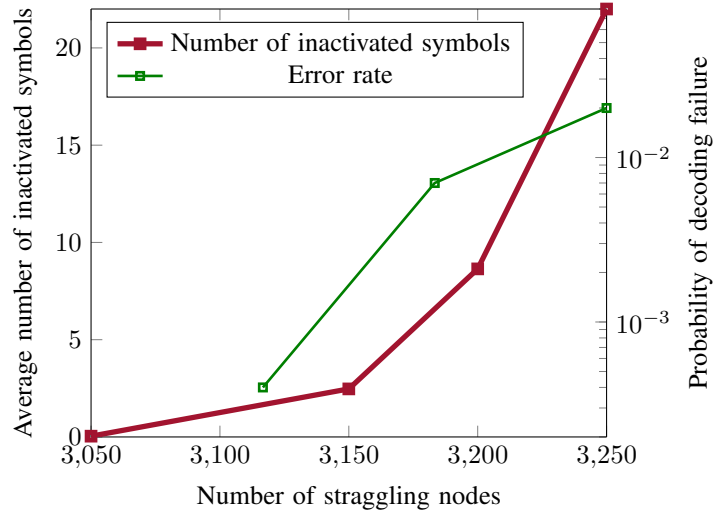

Next, we simulate the performance of the optimal decoder as described in Section~\ref{sec:optimal_decoding}. An $(10000,6400)$ FR code is formed by concatenating an FLT code with a Product code as described above. 
We generate the coefficient matrix $\mathbf{M}$ of the FLT code as mentioned in Section \ref{sec:sysmodel}, i.e., the non-zero entries are chosen to be i.i.d Gaussian random variables.  In Fig.~\ref{fig:LT_optcode_complexity}, we have plotted the number of inactivated symbols that the master node must recover and the probability that the master node is unable to recover the output matrix $\mathbf{C}$ after receiving computations from $N=P-S$ workers.   
\bibliographystyle{IEEEtran}
\bibliography{IEEEabrv,collab}
\end{document}